\documentclass{cimento}

\usepackage{epsfig}

\newcommand{\be}{\begin{equation}}
\newcommand{\ee}{\end{equation}}
\newcommand{\bq}{\begin{eqnarray}}
\newcommand{\eq}{\end{eqnarray}}

\newcommand{\D}{\mathrm{d}}

\newcommand{\I}{\mathrm{i}}

\def\lsim{\mathrel{\rlap{\lower4pt\hbox{\hskip1pt$\sim$}}\raise1pt\hbox{$<$}}}
\def\gsim{\mathrel{\rlap{\lower4pt\hbox{\hskip1pt$\sim$}}\raise1pt\hbox{$>$}}}
\def\Vec#1{\mathpalette{\VVec}{#1}}                  
\def\VVec#1#2{\mbox{\boldmath$#1#2$\unboldmath}}
\def\nostrocostruttino#1\over#2{\mathrel{\mathop{\kern 0pt \rlap
{\hbox{$#1$}}} \hbox{\kern-.135em $#2$}}}

\newlength{\dhatheight}

\def\anti#1{\mathpalette{\@anti}{#1}#1}
\def\@anti#1#2{\sbox0{$#1#2$}
  \makebox[0pt][l]{$#1\kern.30\ht0\overline{\kern-.35\ht0\phantom{#2}}$}}

\title{Spin and transverse momentum dependent Fracture Function in SIDIS}
\author{\underline {A.~Kotzinian}\from{ins:y}\from{ins:b}
\ETC,
M.~Anselmino\from{ins:b}\from{ins:t},
        \atque
V.~Barone\from{ins:a}}
\instlist{
\inst{ins:y}{\it Yerevan Physics Institute, 2 Alikhanyan Brothers St., 375036 Yerevan, Armenia}
\inst{ins:b} {INFN, Sezione di Torino, 10125 Torino, Italy}
\inst{ins:t} {\it Dipartimento di Fisica Teorica, Universit{\`a}
di Torino}

\inst{ins:a}{\it Di.S.T.A., Universit{\`a} del Piemonte
Orientale ``A. Avogadro''; \\
INFN, Gruppo Collegato di Alessandria,  15121 Alessandria, Italy}}

\PACSes{
\PACSit{13.87.Fh}{Fragmentation into hadrons.}
\PACSit{13.88.+e}{Polarization in interactions and scattering.}}

\begin{document}

\maketitle

\begin{abstract}
The recently developed leading twist formalism for spin and transverse-momentum dependent fracture functions is shortly described. We demonstrate that the process of double hadron production in polarized SIDIS -- with one spinless hadron produced in the current fragmentation region (CFR) and another in the target fragmentation region (TFR) -- would  provide access to all 16 leading twist fracture functions. Some particular cases are presented.
\end{abstract}

\section{Introduction}
So far most SIDIS experiments were studied in the CFR, where an adequate theoretical formalism based on distribution and fragmentation functions has been established (see for example Ref.~\cite{Bacchetta:2006tn}). However, for a full understanding of the hadronization process after the hard lepton-quark scattering, also the factorized approach to SIDIS description in the TFR
has to be explored. The corresponding theoretical basis -- the fracture functions formalism -- was established in Ref.~\cite{Trentadue:1993ka}
for hadron transverse momentum integrated unpolarized cross-section. Recently this approach was generalized~\cite{Anselmino:2011ss} to the spin and transverse momentum dependent case (STMD).

We use the standard DIS notations
and in the $\gamma^*-N$ c.m. frame we define the $z$-axis along the direction
of $\Vec q$ (the virtual photon momentum) and the $x$-axis along ${\Vec \ell}_T$,
the  lepton transverse momentum. The kinematics of the produced hadron in the TFR is defined by the variable $\zeta = P_h^-/P^- \simeq E_h/E$ and its transverse momentum
$\Vec P_{h\perp}$ (with magnitude $P_{h\perp}$ and azimuthal angle $\phi_h$). The azimuthal angle of the nucleon transverse polarization is denoted as $\phi_S$.

The STMD fracture functions ${\cal M}$ has a clear probabilistic meaning:
it is the conditional probability to produce a hadron $h$ in the TFR when the
hard scattering occurs on a quark $q$ from the target nucleon $N$.

The most general expression of the LO STMD fracture functions for unpolarized ($\mathcal{M}^{[\gamma^-]}$), longitudinally polarized ($\mathcal{M}^{[\gamma^-\gamma_5]}$) and transversely polarized ($\mathcal{M}^{[\I \, \sigma^{i -} \gamma_5]}$) quarks are introduced in the expansion of the leading twist
projections as~\cite{Anselmino:2011ss,Anselmino:2011bb}:
\begin{eqnarray}
\mathcal{M}^{[\gamma^-]} &=& \hat{u}_1
+ \frac{ {\Vec P}_{h\perp} \times  {\Vec S}_\perp}{m_h} \, \hat{u}_{1T}^h
+ \frac{ {\Vec k}_\perp \times {\Vec S}_\perp}{m_N} \, \hat{u}_{1T}^{\perp}
+ \frac{S_\parallel \, ( {\Vec k}_\perp \times  {\Vec P}_{h\perp})}{m_N \, m_h}
  \, \hat{u}_{1L}^{\perp h} \label{up-frf} \\
\mathcal{M}^{[\gamma^-\gamma_5]} & = &
S_\parallel \, \hat{l}_{1L}
+ \frac{{\Vec P}_{h\perp} \cdot {\Vec S}_\perp}{m_h} \, \hat{l}_{1T}^h
+ \frac{ {\Vec k}_\perp \cdot  {\Vec S}_\perp}{m_N} \, \hat{l}_{1T}^{\perp}
+ \frac{ {\Vec k}_\perp \times {\Vec P}_{h\perp}}{m_N \, m_h} \,
  \hat{l}_1^{\perp h} \label{lp-frf} \\
\mathcal{M}^{[\I \, \sigma^{i -} \gamma_5]} & = & S_\perp^i \, \hat{t}_{1T}
+ \frac{S_\parallel \, P_{h\perp}^i}{m_h} \, \hat{t}_{1L}^h
+ \frac{S_\parallel \, k_\perp^i}{m_N} \, \hat{t}_{1L}^{\perp}
+ \, \frac{( {\Vec P}_{h\perp} \cdot {\Vec S}_\perp)
\, P_{h\perp}^i}{m_h^2} \, \hat{t}_{1T}^{hh}
\nonumber \\ & &
+ \frac{( {\Vec k_\perp} \cdot  {\Vec S}_\perp)
\, k_\perp^i}{m_N^2} \, \hat{t}_{1T}^{\perp \perp}
+ \, \frac{({\Vec k}_\perp \cdot  {\Vec S}_\perp)
\, P_{h\perp}^i - ( {\Vec P}_{h\perp} \cdot {\Vec S}_\perp)
\, k_\perp^i }{m_N m_h} \, \hat{t}_{1T}^{\perp h}
\nonumber \\
& & + \, \frac{\epsilon_{\perp}^{ij} \, P_{h\perp j}}{m_h}
\, \hat{t}_1^h
+ \frac{\epsilon_{\perp}^{ij} \, k_{\perp j}}{m_N}
\, \hat{t}_1^{\perp}\,,
\label{tp-frf}
\end{eqnarray}
where ${\Vec k}_\perp$ is the quark transverse momentum and by the vector
product of two-dimensional vectors ${\bf a}$ and ${\bf b}$ we mean the
pseudo-scalar quantity $ {\bf a} \times  {\bf b} =
\epsilon^{i j} \, a_i b_j =  a   b  \, \sin (\phi_b - \phi_a)$.
All fracture functions depend on the scalar variables $x_B, k_\perp^2, \zeta,
P_{h\perp}^2$ and ${\Vec k}_\perp \cdot {\Vec P}_{h\perp}$.

The single hadron production in the TFR of SIDIS does not provide access to all fracture functions.

\section{Double hadron leptoproduction (DSIDIS)}
In order to have access to all fracture functions one has to "measure" the
scattered quark transverse polarization, for example exploiting he Collins effect~\cite{Collins:1992kk} -- the azimuthal correlation of the fragmenting quark transverse polarization, ${\Vec s}'_T$, with the produced hadron transverse momentum, ${\Vec p}_\perp$:
\be
D(z,{\Vec p}_\perp) = D_1(z, p_\perp^2) + \frac{{\Vec p}_\perp \times
{\Vec s}'_T}{m_h}H_1^\perp(z, p_\perp^2)\, ,
\ee
where $s'_T=D_{nn}(y)\,s_T$ and $\phi_{s'}=\pi-\phi_s$ with
$D_{nn}(y)= [2(1-y)]/[1+(1-y)^2]\>$.

Let us consider a double hadron production process (DSIDIS)
\begin{equation}\label{dsidis}
l(\ell) + N(P) \to l(\ell') + h_1(P_1) + h_2(P_2) + X
\end{equation}
with (unpolarized) hadron 1 produced in the CFR ($x_{F1}>0$) and hadron 2 in the TFR ($x_{F2}<0)$, see Fig.~1.  For hadron $h_1$ we will use the ordinary scaled variable $z_1 = P_1^+/k'^+ \simeq P{\cdot}P_1/P{\cdot}q$ and its transverse
momentum ${\Vec P}_{1\perp}$ (with magnitude $P_{1\perp}$ and azimuthal angle
$\phi_1$) and for hadron $h_2$ the variables $\zeta_2 = P_2^-/P^- \simeq E_2/E$
and ${\Vec P}_{2\perp}$ ($P_{2\perp}$ and $\phi_2$).

\begin{figure}[h!]
\begin{center}
\label{fig:sidis-assoc}
  \includegraphics[height=.15\textheight]{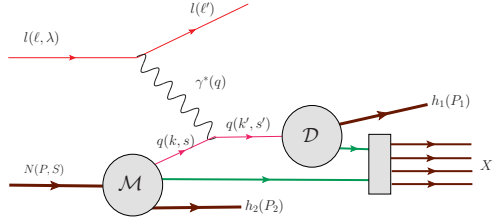}
  \caption{DSIDIS description in factorized approach at LO.}
\end{center}
\end{figure}

The LO expression for the DSIDIS cross-section includes all fracture functions:
\bq \label{cs-2h}
&& \hspace{1.cm}
\frac{\D\sigma^{l(\ell,\lambda)+N(P,S) \to l(\ell')+h_1(P_1)+h_2(P_2)+X}}
{\D x \, \D y \, \D z_1 \, \D\zeta_2 \, \D^2 {\Vec P_{1\perp}} \,
\D^2 {\Vec P_{2\perp}} \, \D \phi_S} =
\frac{\alpha^2\,x_B}{Q^4 \, y}\left[ 1+(1-y)^2 \right] \times
\\ &&
\bigg(\mathcal{M}^{[\gamma^-]}_{h_2} \otimes D_{1q}^{h_1} + \lambda \,
D_{ll}(y)\, \mathcal{M}^{[\gamma^-\gamma_5]}_{h_2} \otimes D_{q}^{h_1}
+ \mathcal{M}^{[\I \, \sigma^{i -} \gamma_5]}_{h_2} \otimes
\frac{{\Vec p}_\perp \times  {\Vec s}'_T}{m_{h_1}}H_{1q}^{\perp h_1} \bigg) =
\nonumber \\
&& \hspace{-0.2cm}
\frac{\alpha^2\,x_B}{Q^4 \, y}\left[ 1+(1-y)^2 \right]
\left(\sigma_{UU} + S_\parallel \,\sigma_{UL} + S_\perp \,\sigma_{UT}+
\lambda \, D_{ll} \, \sigma_{LU} + \lambda \,S_\parallel D_{ll}\,\sigma_{LL}
+\lambda \, S_\perp D_{ll}\,\sigma_{LT} \right)\, ,
\nonumber
\eq
where $D_{ll}(y) = {y(2-y)}/{1+(1-y)^2}$ .

\section{Examples of unintegrated cross-sections: beam spin asymmetry}

We show here explicit expressions only for $\sigma_{UU}$ and $\sigma_{LU}$\footnote{Expressions for other terms are available in~\cite{Kotzinian:DIS2011}.}
\bq\label{s_uu}
\sigma_{UU}  =  F_0^{{\hat u} \cdot D_1}
& - & D_{{nn}} \Bigg[\frac{P_{{1\perp}}^2 }{m_1 m_N}\, F_{{kp1}}^{{\hat t}^\perp \cdot H_1^\perp}\,{\cos}(2 \phi _1)
+ \frac{P_{{1\perp}} P_{{2\perp}} }{m_1 m_2}\, F_{{p1}}^{{\hat t}^h
\cdot H_1^\perp}\, {\cos}(\phi _1+\phi _2)\nonumber \\
& + & \left(\frac{P_{{2\perp}}^2 }{m_1 m_N}\, F_{{kp2}}^{{\hat t}^\perp
\cdot H_1^\perp} + \frac{P_{{2\perp}}^2 }{m_1 m_2}\, F_{{p2}}^{{\hat t^h}
\cdot H_1^\perp}\right)\, {\cos}(2 \phi _2)\Bigg].
\eq
\be
\sigma_{LU} = -\frac{ P_{{1\perp}} P_{{2\perp}}}{m_2 m_N}  F_{{k1}}^{{\hat l}^{\perp h}\cdot D_1} \, \sin(\phi _1-\phi _2)
\, ,
\ee
where the structure functions $F_{...}^{...}$ are specific convolutions~\cite{Kotzinian:DIS2011, Anselmino:2012zz} of fracture and fragmentation functions depending on $x, z_1, \zeta_2, P_{1\perp}^2,  P_{2\perp}^2, {\Vec P}_{1\perp} \cdot
{\Vec P}_{2\perp}$.

We notice the presence of terms similar to the Boer-Mulders term appearing in the usual CFR of SIDIS. What is new in DSIDIS is the LO beam spin SSA, absent in the CFR of SIDIS.
We further notice that the DSIDIS structure functions may depend in principle on the relative azimuthal angle of the two hadrons, due to presence of the last term among their arguments: ${\Vec P}_{1\perp} \cdot {\Vec P}_{2\perp} =
P_{1\perp} P_{2\perp}\cos(\Delta \phi)$ with $\Delta \phi=\phi_1-\phi_2$.
This term arise from ${\Vec k}_\perp \cdot  {\Vec P}_\perp$ correlations in STMD fracture functions and can generate a long range correlation between hadrons produced in CFR and TFR. In practice it is convenient to chose as independent azimuthal angles $\Delta \phi$ and $\phi_2$.

Let us finally consider the beam spin asymmetry defined as
\be
A_{LU}(x, z_1, \zeta_2, P_{1\perp}^2,  P_{2\perp}^2, \Delta \phi) =
\frac{\int \D \phi_2 \, \sigma_{LU}}{\int \D \phi_2 \, \sigma_{UU}}=
\frac{-\frac{ P_{{1\perp}} P_{{2\perp}}}{m_2 m_N}  F_{{k1}}^{{\hat l}^{\perp h}\cdot D_1} \, \sin(\Delta \phi)}{F_0^{{\hat u} \cdot D_1}}\,\cdot
\ee
If one keeps only the linear terms of the corresponding fracture function
expansion in series of ${\Vec P}_{1\perp} \cdot {\Vec P}_{2\perp}$ one obtains
the following azimuthal dependence of DSIDIS beam spin asymmetry:
\be
A_{LU}(x, z_1, \zeta_2, P_{1\perp}^2, P_{2\perp}^2) =
a_1 \sin(\Delta \phi) + a_2 \sin(2\Delta \phi)
\ee
with the amplitudes $a_1,a_2$ independent of azimuthal angles.

In Fig.~2 we present the first preliminary results~\cite{Avakian-Pisano} for $A_{LU}$ asymmetry from CLAS experiment at JLab with $\pi^+$ produced in CFR and $\pi^-$ -- in TFR.  The nonzero effect were observed!

\begin{figure}[h!]
\begin{center}
\label{fig:alu}
  \includegraphics[height=.32\textheight]{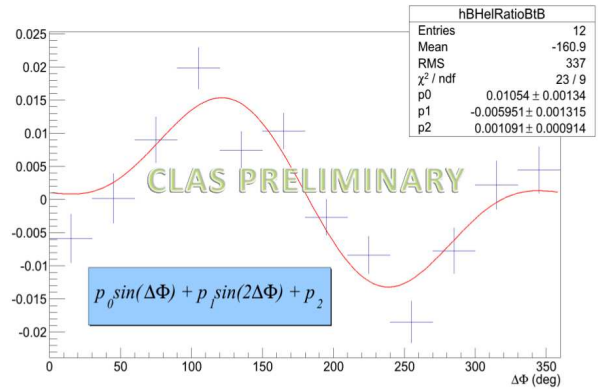}
  \caption{The preliminary results for $A_{LU}$ asymmetry from CLAS experiment at JLab.}
\end{center}
\end{figure}

We stress that the ideal opportunities to test the predictions of the present
approach to DSIDIS, would be the future JLab 12 upgrade, in progress, and the
EIC facilities, in the planning phase.


\begin{thebibliography}{0}

\bibitem{Bacchetta:2006tn}
  A.~Bacchetta, M.~Diehl, K.~Goeke, A.~Metz, P.~J.~Mulders and M.~Schlegel,
  JHEP {\bf 0702}, 093 (2007)
  [arXiv:hep-ph/0611265].

\bibitem{Trentadue:1993ka}
  L.~Trentadue and G.~Veneziano,
  Phys.\ Lett.\  B {\bf 323}, 201 (1994).

\bibitem{Anselmino:2011ss}
  M.~Anselmino, V.~Barone and A.~Kotzinian,
  Phys.\ Lett.\  B {\bf 699}, 108 (2011)
  [arXiv:1102.4214 [hep-ph]].

\bibitem{Anselmino:2011bb}
  M.~Anselmino, V.~Barone and A.~Kotzinian,
  Phys.\ Lett.\ B {\bf 706} (2011) 46
  [arXiv:1109.1132 [hep-ph]].

\bibitem{Collins:1992kk}
  J.~C.~Collins,
  Nucl.\ Phys.\  B {\bf 396}, 161 (1993)
  [arXiv:hep-ph/9208213].

\bibitem{Kotzinian:DIS2011}
A.~Kotzinian, {\it SIDIS in target fragmentation region}, Talk at XIX International Workshop on Deep-Inelastic Scattering and Related Subjects (DIS 2011), April 11-15, 2011, Newport News, VA USA,\\
https://wiki.bnl.gov/conferences/images/3/
3b/Parallel.Spin.AramKotzinian.Thursday14.\\talk.pdf

\bibitem{Anselmino:2012zz}
  M.~Anselmino, V.~Barone and A.~Kotzinian,
  Phys.\ Lett.\ B {\bf 713} (2012) 317.

\bibitem{Avakian-Pisano}
  H. Avakian and S.Pisano, Private communication.

\end{thebibliography}
\end{document}